\documentclass[12pt]{article}

\pdfoutput=1

\usepackage{graphics}
\usepackage{epsfig}

\usepackage{amsfonts}
\usepackage{amssymb}
\usepackage{amsmath}
\usepackage{dsfont}
\usepackage{pifont}
\usepackage{bbm}
\usepackage{multirow}
\usepackage{latexsym}
\usepackage{verbatim}
\usepackage{mcite}
\usepackage{cite}
\usepackage{units}
\usepackage[all]{xy}
\usepackage{cancel}
\usepackage{slashed}
\usepackage{hyperref}

\def\beq{\begin{equation}}
\def\eeq{\end{equation}}
\def\beqa{\begin{eqnarray}}
\def\eeqa{\end{eqnarray}}

\def\a{{\alpha}}
\def\b{{\beta}}

\def\g{{\gamma}}

\def\d{{\delta}}

\def\l{{\lambda}}
\def\m{{\mu}}
\def\n{{\nu}}

\def\bfone{\relax{\rm 1\kern-.35em 1}}

\newcommand{\be}{\begin{equation}}
\newcommand{\ee}{\end{equation}}
\newcommand{\ben}{\begin{displaymath}}
\newcommand{\een}{\end{displaymath}}
\newcommand{\bea}{\begin{eqnarray}}
\newcommand{\eea}{\end{eqnarray}}

\newcommand{\bean}{\begin{eqnarray*}}
\newcommand{\eean}{\end{eqnarray*}}

\DeclareMathAlphabet{\mathpzc}{OT1}{pzc}{m}{it}

\begin{document}
\pagestyle{plain}


\makeatletter \@addtoreset{equation}{section} \makeatother
\renewcommand{\thesection}{\arabic{section}}
\renewcommand{\theequation}{\thesection.\arabic{equation}}
\renewcommand{\thefootnote}{\arabic{footnote}}


\setcounter{page}{1} \setcounter{footnote}{0}


\begin{titlepage}
\begin{flushright}
UUITP-11/15\\
\end{flushright}

\bigskip

\begin{center}

\vskip 0cm

{\LARGE \bf  All gaugings and stable de Sitter in\\[2mm] D=7 half-maximal supergravity} \\[6mm]

\vskip 0.5cm

{\bf Giuseppe Dibitetto$^1$, Jose J. Fern\'andez-Melgarejo$^2$ \,and\, Diego Marqu\'es$^3$}\let\thefootnote\relax\footnote{giuseppe.dibitetto@physics.uu.se, josejuan@physics.harvard.edu, diegomarques@iafe.uba.ar}\\

\vskip 0.5cm

{\em 
$^1$Institutionen f\"or fysik och astronomi, University of Uppsala, \\ Box 803, SE-751 08 Uppsala, Sweden \\
$^2$Jefferson Physical Laboratory, Harvard University, \\ Cambridge, MA 02138, USA \\
$^3$Instituto de Astronom\'{\i}a y F\'{\i}sica del Espacio (CONICET-UBA) \\ C.C. 67 - Suc. 28, 1428 Buenos Aires, Argentina}

\vskip 0.8cm

\end{center}

\vskip 1cm

\begin{center}

{\bf ABSTRACT}\\[3ex]

\begin{minipage}{13cm}
\small

We study the general formulation of gauged supergravity in seven dimensions with sixteen supercharges keeping duality covariance by means of the embedding tensor formalism. We first classify all inequivalent duality orbits of consistent deformations. Secondly, we analyse the complete set of critical points in a systematic way. Interestingly, we find the first examples of stable de Sitter solutions within a theory with such a large amount of supersymmetry. 

\end{minipage}

\end{center}

\vfill

\end{titlepage}


\tableofcontents

\section{Introduction}
\label{sec:introduction}

Supersymmetry preserving flux compactifications of string and M-theory are described by gauged supergravities. The introduction of background fluxes generates a scalar potential in the effective action, providing a successful arena for moduli stabilisation and supersymmetry breaking. The embedding tensor formalism \cite{Nicolai:2000sc,deWit:2002vt} (see also \cite{Samtleben:2008pe} and references therein) provides an efficient formulation of such theories in which all possible flux-deformations are encoded into a single object, the embedding tensor. Much effort has been made in recent years to establish a correspondence between its components and the consistent flux-deformations that arise in string theory compactifications \cite{Aldazabal:2008zza,Dall'Agata:2009gv,Dibitetto:2011gm,Dibitetto:2012ia,Dibitetto:2014sfa}. While some of these deformations (dubbed ``geometric'') can be rapidly identified with metric and $p$-form flux backgrounds, the higher-dimensional origin of others is less clear (and were then named ``non-geometric''). Interestingly, in this formulation the global symmetries inherited form the duality symmetries of the parent theories are manifest. These symmetries mix geometric and non-geometric deformations, allowing to identify the origin of non-geometric fluxes as the result of flux-compactifications in dual backgrounds. 

Regardless of their higher dimensional origin, it is clear that gauged supergravities offer a window to look into the vacuum structure of string theory. It is then of interest to explore and classify their possible gaugings and critical points. This is clearly a very ambitious programme, since the lower dimensional the gauged supergravity, the larger the space of possible deformations allowed by the embedding tensor. Still, it is worth the effort and here we give a small step in this direction. Before we comment on our main results, let us first offer a glimpse into previous results.

Among the three possible types of vacua: AdS, Mkw and dS, the latter is necessarily SUSY breaking and therefore the hardest to find in supersymmetric theories. Since we happen to live in such a vacuum, it is possibly the most interesting one. One finds instead a large landscape of AdS vacua which is mostly of interest in the context of the AdS/CFT correspondence, and there are many known examples of Mkw vacua of special interest in stringy constructions of standard-like models. Stable dS vacua, on the contrary, remain quite reluctant to discovery. We do have some knowledge of why this is the case, specially after some no-go theorems (see for example \cite{Maldacena:2000mw,Hertzberg:2007wc}) and after many studies that relate a positive cosmological constant with instabilities (among which we can mention \cite{Covi:2008ea,Bena:2009xk,Danielsson:2012et}). Although in theories with small amount of supersymmetry there are very interesting examples of dS vacua \cite{deCarlos:2009qm,Danielsson:2012by,Blaback:2013ht}, to our knowledge, there is no known fully stable dS vacuum in (half-)maximal gauged supergravity.

We then find it of interest to focus on particularly simple gauged supergravities in which an exhaustive analysis can be made outside the regions excluded by no-go theorems and previous surveys. Here we consider seven-dimensional half-maximal gauged supergravity \cite{Townsend:1983kk}, which is rich enough in structure to offer an intricate moduli space and at the same time is simple enough to make an exhaustive analysis. We begin by identifying the scalars of the theory \eqref{scalarM1}-\eqref{scalarM2}, the possible gaugings \eqref{Gauge_Gen} with their corresponding quadratic constraints \eqref{QCHalf_Max}, the scalar potential \eqref{VHalf_Max} and the shift matrices \eqref{T-tensor}. The flux-deformations split into three types. On the one hand we have the universal half-maximal gaugings, consisting of the ``three-form'' $f_{ABC}$ and the unimodular deformations $\xi_A$, and on the other there is a (non-gauging) massive deformation $\theta$. We then classify all orbits of solutions to the quadratic constraints (tables \ref{table:orbits1} and \ref{table:orbits2}), therefore finding all consistent seven-dimensional gauged supergravities with sixteen supercharges. The space of deformations splits into two branches, one with $\theta = 0$ ({\bf branch 1}) and another one with $\xi_A = 0$ ({\bf branch 2}). While the classification of orbits of {\bf branch 2} was exhaustively performed in \cite{Dibitetto:2012rk}, here we complete the classification for {\bf branch 1} by including the unimodular deformations $\xi_A$.

We then move to the analysis of critical points, following the going-to-the-origin approach \cite{Dibitetto:2011gm}. While {\bf branch 1} contains only Mkw vacua, {\bf branch 2} allows for non-semisimple configurations that exhibit both AdS and Mkw, and semisimple configurations with a large variety of minima. Interestingly, we find an $\textrm{SO}(1,3)$ gauging configuration with an AdS-Mkw-dS transition vacuum containing a fully stable dS window. We believe this is the first example of stable dS vacua in half-maximal supergravity.

Let us briefly sketch the structure of the paper. In section 2 we introduce the theory and all the elements that are necessary to explore its moduli space. Section 3 is devoted to classify all the duality orbits of deformations. In section 4 we perform the analysis of critical points and conclude with a discussion in section 5. For completion we also include an appendix where we collected some technical material that may be relevant for our analysis.

%

\section{The half-maximal $D=7$ gauged supergravities}
\label{sec:7DSugra}

Half-maximal (ungauged) supergravity in seven dimensions coupled to three vector multiplets can be obtained by reducing type I supergravity in ten dimensions on a $\mathbb{T}^{3}$. 
The theory possesses $16$ supercharges which can be rearranged into a pair of symplectic-Majorana (SM) spinors transforming as a doublet of $\textrm{SU}(2)_{R}$. The full Lagrangian enjoys a global 
symmetry given by
\be
G_{0} \ = \ \mathbb{R}^{+}_{\Sigma} \, \times \, \textrm{SO}(3,3) \ \approx \ \mathbb{R}^{+}_{\Sigma} \, \times \, \textrm{SL}(4) \ .
\notag
\ee
The $(64_{B} \ + \ 64_{F})$ bosonic and fermionic propagating degrees of freedom (dof's) of the theory are then rearranged into irrep's of $G_{0}$ as described in table~\ref{Table:dofs}. We refer to 
the appendix for a summary of our notations for all different indices used throughout the paper.
\begin{table}[h!]
\renewcommand{\arraystretch}{1}
\begin{center}
\scalebox{1}[1]{
\begin{tabular}{|c|c|c|c|c|}
\hline
fields & $\textrm{SO}(5)$ irrep's & $\mathbb{R}^{+} \, \times \, \textrm{SL}(4)$ irrep's & $\textrm{SU}(2)_{R} \, \times \, \textrm{SU}(2)$ irrep's & \# dof's  \\
\hline \hline
${e_{\mu}}^{a}$ & $\textbf{14}$ & $\textbf{1}_{(0)}$ & $(\textbf{1},\textbf{1})$ & $14$ \\
\hline
${A_{\mu}}^{[mn]}$ & $\textbf{5}$ & $\textbf{6}_{(+1)}$ & $(\textbf{1},\textbf{1})$ & $30$ \\
\hline
$B_{\m\n}$ & $\textbf{10}$ & $\textbf{1}_{(+2)}$ & $(\textbf{1},\textbf{1})$ & $10$ \\
\hline
$\Sigma$ & $\textbf{1}$ & $\textbf{1}_{(+1)}$ & $(\textbf{1},\textbf{1})$ & $1$ \\
\hline
${{\mathcal{V}}_{m}}^{\a\hat{\a}}$ & $\textbf{1}$ & $\textbf{4}^{\prime}_{(0)}$ & $(\textbf{2},\textbf{2})$ & $9$ \\
\hline
\hline
$\psi_{\m\a}$ & $\textbf{16}$ & $\textbf{1}_{(0)}$ & $(\textbf{2},\textbf{1})$ & $32$ \\
\hline
${\chi}_{\a}$ & $\textbf{4}$ & $\textbf{1}_{(0)}$ & $(\textbf{2},\textbf{1})$ & $8$ \\
\hline
$\lambda^{\a\hat{\a}\hat{\b}}$ & $\textbf{4}$ & $\textbf{1}_{(0)}$ & $(\textbf{2},\textbf{3})$ & $24$ \\ 
\hline
\end{tabular}
}
\end{center}
\caption{{\it The on-shell field content of (ungauged) half-maximal supergravity in $D=7$. Each field is massless and hence transforms in some irrep of the corresponding little group $\textrm{SO}(5)$ w.r.t.
spacetime diffeomorphisms and local Lorentz transformations. Please note that, in the $\textrm{SL}(4)$ scalar coset representative ${{\mathcal{V}}_{m}}^{\a\hat{\a}}$, one needs to subtract the number
of unphysical scalars corresponding with $\textrm{SO}(4)$ generators in order to come up with the correct number of dof's, \emph{i.e.} $9$.}} \label{Table:dofs}
\end{table}

As one can see from table~\ref{Table:dofs}, the scalar sector of the theory contains an $\mathbb{R}^{+}$ scalar denoted by $\Sigma$ and an $\frac{\textrm{SO}(3,3)}{\textrm{SO}(3)\times\textrm{SO}(3)}$ 
coset representative denoted by $\mathcal{M}_{AB}$. However, by exploiting the isomorphism between $\textrm{SO}(3,3)$ and $\textrm{SL}(4)$ at the level of their Lie algebras, it is particularly convenient
to parametrise this set of scalars by an $\frac{\textrm{SL}(4)}{\textrm{SO}(4)}$ coset representative which we denote by $M_{mn}$. In terms of the vielbein ${{\mathcal{V}}_{m}}^{\a\hat{\a}}$ appearing
in table~\ref{Table:dofs}, $M_{mn}$ can be constructed as 
\be
M_{mn} \ = \ {{\mathcal{V}}_{m}}^{\a\hat{\a}} \, {{\mathcal{V}}_{n}}^{\b\hat{\b}} \, \epsilon_{\a\b} \, \epsilon_{\hat{\a}\hat{\b}} \ ,
\label{scalarM1}
\ee
where $\epsilon_{\a\b} \, \epsilon_{\hat{\a}\hat{\b}}$ can be viewed as the invariant metric of $\textrm{SU}(2)_{R} \, \times \, \textrm{SU}(2)$, which can be brought into the form of an $\mathds{1}_{4}$.
Given a realisation of $M_{mn}$, $\mathcal{M}_{AB}$ can then be obtained as
\be
\mathcal{M}_{AB} \ = \ \frac{1}{2} \, [G_{A}]^{mp}[G_{B}]^{nq}\,M_{mn}M_{pq} \ ,
\label{scalarM2}
\ee
in terms of the 't Hooft symbols $[G_{A}]^{mn}$ introduced in appendix~\ref{App:'tHooft}. 

The kinetic Lagrangian for the scalar sector reads
\be
\label{Lkin}
\mathcal{L}_{\textrm{kin}} \ = \ -\frac{5}{2}\,\Sigma^{-2}\,\left(\partial\Sigma\right)^{2} \ + \ \frac{1}{16}\,\partial_{\m}\mathcal{M}_{AB}\,\partial^{\m}\mathcal{M}^{AB} \ = \
-\frac{5}{2}\,\Sigma^{-2}\,\left(\partial\Sigma\right)^{2} \ + \ \frac{1}{8}\,\partial_{\m}M_{mn}\,\partial^{\m}M^{mn} \ ,
\ee
where $\mathcal{M}^{AB}$ and $M^{mn}$ denote the inverse of $\mathcal{M}_{AB}$ and $M_{mn}$, respectively.

As a consequence of the linear constraint (LC), the deformations of the theory described by a generalised embedding tensor need to transform in the following $G_{0}$ irrep's \cite{Dibitetto:2012rk}
\be
\begin{array}{lclclclclc}
\Theta & \in & \underbrace{\textbf{1}_{(-4)}}_{\theta} & \oplus & \underbrace{ \textbf{10}^{\prime}_{(+1)}}_{Q_{(mn)}} & \oplus & \underbrace{ \textbf{10}_{(+1)}}_{{\tilde{Q}}^{(mn)}} & \oplus & 
\underbrace{\textbf{6}_{(+1)}}_{\xi_{[mn]}} & ,
\end{array}
\notag
\ee 
where $\theta$ can be viewed as a St\"uckelberg coupling defining as a so-called $p=3$-type deformation \cite{Bergshoeff:2007vb}, whereas all the other irreducible pieces correspond to gaugings. 
In particular, $Q$ \& $\tilde{Q}$ can be used in order to gauge a subgroup of $\textrm{SL}(4)$, whereas $\xi$ necessarily gauges the $\mathbb{R}^{+}_{\Sigma}$ generator as well as a suitable subgroup of
$\textrm{SL}(4)$. 

It is worth mentioning that the 't Hooft symbols given in appendix~\ref{App:'tHooft} may be used to map $Q \ \oplus \ \tilde{Q}$ and $\xi$ into a 3-form $f_{ABC}$ (self-dual (SD) and
anti-self-dual (ASD) part) and a vector $\xi_{A}$ of $\textrm{SO}(3,3)$, respectively. Such $f$ and $\xi$ characterise the universal sector of consistent gaugings of half-maximal theories which exist 
in any dimension in the presence of vector multiplets.

The generators of the gauge algebra can be written as
\be
\label{Gauge_Gen}
{\left(X_{mn}\right)_{pq}}^{rs} \ = \ \frac{1}{2}\,\delta^{[r}_{[m}\,Q_{n][p}\,\delta^{s]}_{q]} \ + \ \frac{1}{4}\,\epsilon_{tmn[p}\,(\tilde{Q}\,+\,\xi)^{t[r}\,\delta^{s]}_{q]} \ ,
\ee
in terms of the embedding tensor. Please note that the ${\left(X_{mn}\right)_{pq}}^{rs}$'s are in general not traceless. In particular their trace is proportional to $\xi$, thus implying that one needs
an embedding tensor in the $\textbf{6}$ in order to gauge the $\mathbb{R}^{+}$ generator outside of $\textrm{SL}(4)$.

The closure of the gauge algebra and more general bosonic consistency impose the following quadratic constraints (QC) on the various irreducible components of the embedding tensor
\be
\label{QCHalf_Max}
\begin{array}{rclc}
\left(\tilde{Q}^{mp}\,+\,\xi^{mp}\right)\,Q_{pn} \ - \ \frac{1}{4}\,\left(\tilde{Q}^{pq}\,Q_{pq}\right)\,\delta^{m}_{n} & = & 0 & , \\[2mm]
Q_{mp}\,\xi^{pn} \ + \ \,\xi_{mp}\,\tilde{Q}^{pn} & = & 0 & , \\[2mm]
\xi_{mn}\,\xi^{mn} & = & 0 & , \\[2mm]
\theta\,\xi_{mn} & = & 0 & , \\[2mm]
\end{array}
\ee
where $\xi^{mn}\,\equiv\,\frac{1}{2}\,\epsilon^{mnpq}\,\xi_{pq}$. The above QC contain irreducible pieces transforming in the $\textbf{1}_{(+2)}\,\oplus\,\textbf{6}_{(-3)}\,\oplus\,
\textbf{15}_{(+2)}$ of $\mathbb{R}^{+}_{\Sigma} \, \times \, \textrm{SL}(4)$.

Gauge invariance and supersymmetry force the scalar potential of the theory to be of the form\footnote{We hereby correct a typo in ref.~\cite{Danielsson:2013qfa} concerning the sign of the $\theta\,\tilde{Q}$ term, 
which, though, does not affect any of the results obtained there.}
\be
\label{VHalf_Max}
\begin{array}{lclc}
V & = & \dfrac{g^{2}}{64}\,\bigg(\theta^{2}\,\Sigma^{8} \ + \ \dfrac{1}{4}\,Q_{mn}Q_{pq}\,\Sigma^{-2}\,\left(2M^{mp}M^{nq}\,-\,M^{mn}\,M^{pq}\right) \ + & \\[1mm]
& + & \dfrac{3}{2}\,\xi_{mn}\xi_{pq}\,\Sigma^{-2}\,M^{mp}M^{nq} \ + \ \dfrac{1}{4}\,\tilde{Q}^{mn}\tilde{Q}^{pq}\,\Sigma^{-2}\,\left(2M_{mp}M_{nq}\,-\,M_{mn}\,M_{pq}\right) \ + & \\[1mm]
& - & \theta\,\left(Q_{mn}M^{mn}\,-\,\tilde{Q}^{mn}M_{mn}\right)\,\Sigma^{3} \ + \ Q_{mn}\tilde{Q}^{mn}\,\Sigma^{-2} \bigg) & ,
\end{array}
\ee
where $g$ denotes an arbitrary gauge coupling\footnote{In section~\ref{sec:critical_points}, we will set $g\,=\,8$ when analysing the set of critical points of the various gauged theories}. The above expression generalises the one given in ref.~\cite{Danielsson:2013qfa} to the case with $\xi_{mn}\neq 0$.

The gauging procedure induces mass terms for the fermions proportional to the gauge coupling constant of the following form
\be
e^{-1}\,\mathcal{L}_{\textrm{f.~mass}} \ \supset \ g\,\left({A_{1}}^{\a\b}\,\bar{\psi}_{\m\a}\,\gamma^{\m\n}\,{\psi}_{\n\b} \ \oplus \ {A_{2}}^{\a\b}\,\bar{\psi}_{\m\a}\,\gamma^{\m}\,{\chi}_{\b} \ \oplus \ {A_{3\,\hat{\a}\hat{\b}\b}}^{\a}\,\bar{\psi}_{\m\a}\,\gamma^{\m}\,{\lambda}^{\b\hat{\a}\hat{\b}}\right) \ ,
\ee
where the shift matrices ${A_{1}}^{\a\b}\,=\,{A_{1}}^{[\a\b]}$, ${A_{2}}^{\a\b}$ and ${A_{3\,\hat{\a}\hat{\b}\b}}^{\a}\,=\,{A_{3\,(\hat{\a}\hat{\b})\b}}^{\a}\,$ are the irreducible components of the T-tensor. These can be written in terms of the embedding tensor as
\be
\label{T-tensor}
\begin{array}{lclc}
{A_{1}}^{\a\b} & = & \frac{1}{8}\,\left(\Sigma^{4}\,\theta\,\epsilon^{\a\b}\,+\,\Sigma^{-1}\,Q_{mn}\,{\mathcal{V}_{\g\hat{\a}}}^{m}{\mathcal{V}_{\d\hat{\b}}}^{n}\,\epsilon^{\g\a}\epsilon^{\d\b}\epsilon^{\hat{\a}\hat{\b}}\,-\,\Sigma^{-1}\,\tilde{Q}^{mn}\,{\mathcal{V}_{m}}^{\a\hat{\a}}{\mathcal{V}_{n}}^{\b\hat{\b}}\,\epsilon_{\hat{\a}\hat{\b}}\right) & , \\[4mm]
{A_{2}}^{\a\b} & = & \frac{1}{8}\,\left(\Sigma^{4}\,\theta\,\epsilon^{\a\b}\,-\,\frac{1}{4}\Sigma^{-1}\,Q_{mn}\,{\mathcal{V}_{\g\hat{\a}}}^{m}{\mathcal{V}_{\d\hat{\b}}}^{n}\,\epsilon^{\g\a}\epsilon^{\d\b}\epsilon^{\hat{\a}\hat{\b}}\,+\,\frac{1}{4}\Sigma^{-1}\,\tilde{Q}^{mn}\,{\mathcal{V}_{m}}^{\a\hat{\a}}{\mathcal{V}_{n}}^{\b\hat{\b}}\,\epsilon_{\hat{\a}\hat{\b}}\right) &   \\[2mm]
& & + \frac{\sqrt{15}}{32}\,\Sigma^{-1}\,\xi_{mn}\,{\mathcal{V}_{\g\hat{\a}}}^{m}{\mathcal{V}_{\d\hat{\b}}}^{n}\,\epsilon^{\g\a}\epsilon^{\d\b}\epsilon^{\hat{\a}\hat{\b}} & , \\[4mm]
{A_{3\,\hat{\a}\hat{\b}\g}}^{\d} & = & \frac{1}{8}\,\Sigma^{-1}\,\left(\,Q_{mn}\,{\mathcal{V}_{\g\hat{\a}}}^{m}{\mathcal{V}_{\b\hat{\b}}}^{n}\,\epsilon^{\b\d}\,-\,\tilde{Q}^{mn}\,{\mathcal{V}_{m}}^{\a\hat{\g}}{\mathcal{V}_{n}}^{\d\hat{\d}}\,\epsilon_{\hat{\g}\hat{\a}}\epsilon_{\hat{\d}\hat{\b}}\epsilon_{\a\g}\right. & \\[2mm]
& & \left.-\frac{\sqrt{3}}{2}\,\xi_{mn}\,{\mathcal{V}_{\a\hat{\a}}}^{m}{\mathcal{V}_{\b\hat{\b}}}^{n}\,\epsilon^{\a\b}\d_{\g}^{\d}\right) & .
\end{array}
\ee

\noindent The conditions for preserving supersymmetry read
\be
{A_{1}}^{\a\b}\,q_{\b} \ \overset{!}{=} \ \sqrt{-\frac{10\,V}{3}}\,q^{\a} \ ,
\ee
for a pair of SM spinors $q_{\a}$ transforming in the fundamental representation of $\textrm{SU}(2)$ or, equivalently,
\be
\begin{array}{lcllc}
{A_{2}}^{\a\b}\,q_{\b} \ \overset{!}{=} \ 0 & , & \textrm{ and } & {A_{3\,\hat{\a}\hat{\b}\g}}^{\d}\,q_{\d} \ \overset{!}{=} \ 0 & .
\end{array}
\ee

In terms of the shift matrices defined in \eqref{T-tensor}, the scalar potential in \eqref{VHalf_Max} can be rewritten as
\be
V \ = \ g^{2} \ \left( - \ \frac{3}{10}\left|A_{1}\right|^{2} \ + \ \frac{4}{5}\left|A_{2}\right|^{2} \ + \ \frac{1}{2}\left|A_{3}\right|^{2} \right) \ ,
\ee
where\footnote{$\textrm{SU}(2)$ indices are raised and lowered by means of $\epsilon_{\a\b}$ and $\epsilon_{\hat{\a}\hat{\b}}$.} $\left|A_{1}\right|^{2}\,\equiv\,{A_{1}}^{\a\b}\,{A_{1}}_{\a\b}$, $\left|A_{2}\right|^{2}\,\equiv\,{A_{2}}^{\a\b}\,{A_{2}}_{\a\b}$ and $\left|A_{3}\right|^{2}\,\equiv\,{A_{3\,\hat{\a}\hat{\b}\g}}^{\d}\,{{{A_{3}}^{\hat{\a}\hat{\b}\g}}}_{\d}$.

Please note that, in the origin of moduli space, the vielbeins of $\frac{\textrm{SL}(4)}{\textrm{SO}(4)}$ read
\be
\left.{\mathcal{V}_{m}}^{\a\hat{\b}}\right|_{\textrm{origin}} \ = \ \frac{1}{\sqrt{2}} \, \left[\Gamma_{\underline{m}}\right]^{\a\hat{\b}} \ ,
\ee
where $\left[\Gamma_{\underline{m}}\right]^{\a\hat{\b}}$ denote the Dirac matrices of $\textrm{SO}(4)$ in the Weyl representation (see appendix~ \ref{App:Gamma} for more details).

\subsection*{$\mathcal{N}=4$, $D=4$ gaugings with 7D origin}

Half-maximal supergravity in $D=4$ with six vector multiplets exhibits manifest $\textrm{SL}(2)\times\textrm{SO}(6,6)$ global symmetry
and it admits embedding tensor deformations which are restricted to transform in the
\be
\begin{array}{lclclc}
\Theta & \in & \underbrace{ (\textbf{2},\textbf{12})}_{\xi_{a \mathbb{M}}} & \oplus & \underbrace{ (\textbf{2},\textbf{220})}_{f_{a [\mathbb{M}\mathbb{N}\mathbb{P}]}} & ,
\end{array}
\notag
\ee
of $\textrm{SL}(2)\times\textrm{SO}(6,6)$ by the LC \cite{Schon:2006kz}.

By reducing a consistent gauged theory in 7D on a $\mathbb{T}^{3}$, one obtains a particular class of embedding tensors in 4D. Hence, in order to identify the parts of $f_{a [\mathbb{M}\mathbb{N}\mathbb{P}]}$ \& $\xi_{a \mathbb{M}}$ which have a seven-dimensional origin, one needs to branch the $\mathcal{N}=4$ embedding tensor w.r.t. our global 7D symmetry $G_0$. This is done through the following chain
\be
\begin{array}{lclclc}
\textrm{SL}(2)_{\Sigma}\times\textrm{SO}(6,6) & \supset & \textrm{SL}(2)_{\Sigma}\times\textrm{SO}(3,3)_{A}\times\textrm{SO}(3,3)_{\hat{A}} & \supset & \mathbb{R}^{+}_{\Sigma}\times\textrm{SO}(3,3)_{A} & ,
\end{array}
\notag
\ee
where $\mathbb{R}^{+}_{\Sigma}$ is a combination of the $\mathbb{R}^{+}$ sitting inside $\textrm{SL}(2)_{\Sigma}$ and one of the Cartan generators of $\textrm{SO}(3,3)_{\hat{A}}$ in the last step one is allowed to identify the fundamental representation of $\textrm{SO}(3,3)_{A}$ with the two-form of $\textrm{SL}(4)$ by using the mapping in appendix~\ref{App:'tHooft}.

The fundamental index $\mathbb{M}$ of $\textrm{SO}(6,6)$ splits as
\be
\begin{array}{lclc}
\mathbb{M} & \longrightarrow & A \ \oplus \ \left(i,\,j,\,k;\,\bar{i},\,\bar{j},\,\bar{k}\right) & ,
\end{array} \notag
\ee
in light-cone coordinates.

The ten scalars of the 7D theory are embedded as
\be
\begin{array}{lcc}
M_{ab} \ = \ \left(\begin{array}{c|c}\Sigma^{2} & \\[2mm]\hline  & \Sigma^{-2}\end{array}\right) & , & \mathcal{M}_{\mathbb{M}\mathbb{N}} \ = \ \left(\begin{array}{c|c}\mathcal{M}_{AB} & \\[2mm]\hline & \begin{array}{c|c}\Sigma^{2}\,\mathds{1}_{3} & \\[2mm]\hline  & \Sigma^{-2}\,\mathds{1}_{3}\end{array}\end{array}\right)
\end{array} 
\ee
into the $\frac{\textrm{SL}(2)}{\textrm{SO}(2)}$ and $\frac{\textrm{SO}(6,6)}{\textrm{SO}(6)\times\textrm{SO}(6)}$ coset representatives, respectively.

The embedding tensor of the 7D theory is embedded into the objects $f_{a [\mathbb{M}\mathbb{N}\mathbb{P}]}$ \& $\xi_{a \mathbb{M}}$ as follows
\be
\begin{array}{lclcclc}
f_{+ ABC} \ = \ f_{ABC} & , &  f_{- \bar{i}\bar{j}\bar{k}} \ = \ \theta & , & \textrm{and} & \xi_{+ A} \ = \ \frac{1}{\sqrt{2}}\,\xi_{A} & ,
\end{array} 
\ee
where 
\be
\begin{array}{lc}
f_{ABC} \ = \ {\left(X_{mn}\right)_{pq}}^{rs} \, \left[G_{A}\right]^{mn} \, \left[G_{B}\right]^{pq} \, \left[G_{C}\right]_{rs} & , \\[2mm]
\xi_{A} \ = \ \xi_{mn} \, \left[G_{A}\right]^{mn} & ,
\end{array} 
\ee
in terms of the objects in \eqref{Gauge_Gen} and the 't Hooft symbols defined in \eqref{tHooft}.

\section{Orbit classification of deformations}
\label{sec:orbits}

Each solution to the QC in \eqref{QCHalf_Max} identifies a consistent deformation of half-maximal 7D supergravity.
The global symmetry group $\mathbb{R}^{+}\times\textrm{SL}(4)\,\approx\,\mathbb{R}^{+}\times\textrm{SO}(3,3)\,$ of the theory can be interpreted as T-duality and, since the QC are manifestly covariant
w.r.t. such global symmetry, the space of solutions is naturally split into duality orbits.

This section is the natural generalisation of the analysis carried out in ref.~\cite{Dibitetto:2012rk} in the case $\theta\,=\,\xi_{mn}\,=\,0$. Due to the last condition in \eqref{QCHalf_Max}, the set of all consistent gaugings is naturally split into two independent branches:
\be
\begin{array}{lccccclc}
\bullet \textbf{ branch 1: }  \qquad\theta \, = \, 0   & , & & & & &
\bullet \textbf{ branch 2: }  \qquad \xi_{mn} \, = \, 0 & .
\end{array}\notag
\ee

\subsection*{Orbits of deformations in branch 1}

When $\theta\,=\,0$, the corresponding deformation can be understood as a gauging in the traditional sense, \emph{i.e.} it is obtained by promoting a suitable subgroup of $G_{0}$ to a local symmetry.
In particular, as we have already observed earlier, $Q$ \& $\tilde{Q}$ purely gauge generators within the $\textrm{SL}(4)$ factor, whereas $\xi$ necessarily gauges the extra $\mathbb{R}^{+}_{\Sigma}$ as
well as a subgroup of $\textrm{SL}(4)$.

The T-duality orbit classifications for gaugings in this branch is presented in table~\ref{table:orbits1}.
\begin{table}[h!]
\begin{center}
\scalebox{0.94}[0.94]{
\begin{tabular}{| c | c | c | c | c |}
\hline
\textrm{ID} & $\xi_{mn}$ & $Q_{mn}/\,\cos\alpha\,$ & $\tilde{Q}^{mn}/\,\sin\alpha\,$ & gauging \\[1mm]
\hline \hline
1 & \multirow{4}{*}{${0}_{4}$} & $\mathds{1}_{4}$ & $\mathds{1}_{4}$  & 
$\begin{array}{ll}\textrm{SO}(4)\ , & \a\,\neq\,\frac{\pi}{4}\\[1mm]\textrm{SO}(3)\ , & \a\,=\,\frac{\pi}{4}\end{array}$\\[1mm]
\cline{1-1}\cline{3-4}\cline{5-5} 2 &  & $\textrm{diag}(1,1,1,-1)$ & $\textrm{diag}(1,1,1,-1)$ & $\textrm{SO}(3,1)$\\[1mm]
\cline{1-1}\cline{3-4}\cline{5-5} 3 &  & $\textrm{diag}(1,1,-1,-1)$ & $\textrm{diag}(1,1,-1,-1)$ & 
$\begin{array}{ll}\textrm{SO}(2,2)\ , & \a\,\neq\,\frac{\pi}{4}\\[1mm]\textrm{SO}(2,1)\ , & \a\,=\,\frac{\pi}{4}\end{array}$\\[1mm]
\hline
\hline
4 & \multirow{2}{*}{${0}_{4}$} & $\textrm{diag}(1,1,1,0)$ & \multirow{2}{*}{$\textrm{diag}(0,0,0,1)$} & $\textrm{CSO}(3,0,1)$ \\[1mm]
\cline{1-1}\cline{3-3}\cline{5-5} 5 &  & $\textrm{diag}(1,1,-1,0)$ & &  $\textrm{CSO}(2,1,1)$\\[1mm]
\hline
\hline
6 & \multirow{5}{*}{$\xi_{0}\,\left(\begin{array}{c|c} \epsilon_{2} & \\ \hline & 0_{2} \end{array}\right)$} & 
\multirow{5}{*}{$\textrm{diag}(1,1,0,0)$} & $\textrm{diag}(0,0,1,1)$ & 
$\begin{array}{ll}\textrm{CSO}(2,0,2)\ , & \left|\xi_0\right|<1\\[1mm]\mathfrak{f}_{1}\quad(\textrm{Solv}_{6})^{*} \ , & \left|\xi_0\right|=1\end{array}$ \\[1mm]
\cline{1-1}\cline{4-4}\cline{5-5} 7 &  &  & $\textrm{diag}(0,0,1,-1)$ & 
$\begin{array}{ll}\textrm{CSO}(2,0,2)\ , & \left|\xi_0\right|<\sqrt{\cos(2\a)}\\[1mm]\textrm{CSO}(1,1,2)\ , & \left|\xi_0\right|>\sqrt{\cos(2\a)}\\[1mm]\mathfrak{g}_{0}\quad(\textrm{Solv}_{6})^{*} \ , & \left|\xi_0\right|=\sqrt{\cos(2\a)}\end{array}$ \\[1mm]
\cline{1-1}\cline{4-4}\cline{5-5} 8 &  &  & $\textrm{diag}(0,0,0,1)$ & $\mathfrak{h}_{1}\quad(\textrm{Solv}_{6})^{*}$ \\[1mm]
\hline
9 & \multirow{4}{*}{$\xi_{0}\,\left(\begin{array}{c|c} \epsilon_{2} & \\ \hline & 0_{2} \end{array}\right)$} & 
\multirow{4}{*}{$\textrm{diag}(1,-1,0,0)$} & $\textrm{diag}(0,0,1,1)$ & $\mathfrak{f}_{2}\quad(\textrm{Solv}_{6})^{*}$ \\[1mm]
\cline{1-1}\cline{4-4}\cline{5-5} 10 &  &  & $\textrm{diag}(0,0,1,-1)$ & $\textrm{CSO}(1,1,2)$ \\[1mm]
\cline{1-1}\cline{4-4}\cline{5-5} 11 &  &  & $\textrm{diag}(0,0,0,1)$ & $\mathfrak{h}_{2}\quad(\textrm{Solv}_{6})^{*}$ \\[1mm]
\hline\hline
12 & $\xi_{0}\,\left(\begin{array}{c|c} \epsilon_{2} & \\ \hline & 0_{2} \end{array}\right)$ & 
$\textrm{diag}(1,0,0,0)$ & $\textrm{diag}(0,0,0,1)$ & $\begin{array}{ll}\mathfrak{l}\,\,\,\,(\textrm{Nil}_{6}(3))^{*}\ , & \xi_0\,\neq\,0\\[1mm]\textrm{CSO}(1,0,3) \ , & \xi_0\,=\,0\end{array}$ \\[1mm]
\hline\hline
13 & $\xi_{0}\,\left(\begin{array}{c|c} \epsilon_{2} & \\ \hline & 0_{2} \end{array}\right)$ & $0_{4}$ & $0_{4}$ & $\left(\mathbb{R}^{+}\ltimes\left(\mathbb{R}^{+}\right)^{3}\right)\times\textrm{U}(1)^{2}$ \\[1mm]
\hline
\end{tabular}
}
\end{center}
\caption{{\it All the T-duality orbits of consistent gaugings in the ${\bf 6} \, \oplus \, {\bf 10} \, \oplus \, {\bf 10^{\prime}}$ of half-maximal supergravity in $D=7$. 
Any value of $(\alpha,\,\xi_0)$ parameterises inequivalent orbits; the range of $\alpha$ is everywhere $-\frac{\pi}{4}\,<\,\alpha\,\le\,\frac{\pi}{4}$, while that of $\xi_0$ is $-1\,\le\,\xi_0\,\le\,1$. 
The shorthand $\epsilon_{2}$ denotes the 2D Levi-Civita symbol. Note that, whenever $\xi_0\neq 0$, one Abelian gauge generator needs to coincide with $\mathbb{R}^{+}_{\Sigma}$. 
For more details on algebras marked with *, see appendix~\ref{App:algebras}.} 
\label{table:orbits1}}
\end{table}

As far as the higher-dimensional origin of the orbits in this branch is concerned, they can all be regarded as generalised twisted reductions of heterotic supergravity and hence they should all be accessible by means of twisted reductions of Double Field Theory (DFT) \cite{Hohm:2010pp}. However, as already noted\footnote{Please note that such non-geometric backgrounds were also found in ref.~\cite{Condeescu:2013yma} to arise from asymmetric
orbifold constructions.} in \cite{Dibitetto:2012rk}, this will generically require a relaxation of the section condition as originally proposed in ref.~\cite{Grana:2012rr}.

\subsection*{Orbits of deformations in branch 2}

When $\xi_{mn}\,=\,0$, the most general consistent deformation of half-maximal 7D supergravity is a combination of a \emph{massive} deformation induced by $\theta$ and a gauging of an arbitrary (up to
six-dimensional) subgroup of $\textrm{SL}(4)$. The consistency conditions and therefore the resulting gauge algebras turn out to be identical to those in the $\theta\,=\,\xi_{mn}\,=\,0$ case
already analysed in detail in ref.~\cite{Dibitetto:2012rk}. We collect in table~\ref{table:orbits2} the corresponding results suitably combined with a non-zero $\theta$.
\begin{table}[h!]
\begin{center}
\scalebox{1}[1]{
\begin{tabular}{| c | c | c | c | c |}
\hline
\textrm{ID} & $\theta$ & $Q_{mn}/\,\cos\alpha\,$ & $\tilde{Q}^{mn}/\,\sin\alpha\,$ & gauging \\[1mm]
\hline \hline
1 & \multirow{4}{*}{$\kappa$} & $\mathds{1}_{4}$ & $\mathds{1}_{4}$  & 
$\begin{array}{ll}\textrm{SO}(4)\ , & \a\,\neq\,\frac{\pi}{4}\\[1mm]\textrm{SO}(3)\ , & \a\,=\,\frac{\pi}{4}\end{array}$\\[1mm]
\cline{1-1}\cline{3-4}\cline{5-5} 2 &  & $\textrm{diag}(1,1,1,-1)$ & $\textrm{diag}(1,1,1,-1)$ & $\textrm{SO}(3,1)$\\[1mm]
\cline{1-1}\cline{3-4}\cline{5-5} 3 &  & $\textrm{diag}(1,1,-1,-1)$ & $\textrm{diag}(1,1,-1,-1)$ & 
$\begin{array}{ll}\textrm{SO}(2,2)\ , & \a\,\neq\,\frac{\pi}{4}\\[1mm]\textrm{SO}(2,1)\ , & \a\,=\,\frac{\pi}{4}\end{array}$\\[1mm]
\hline
\hline
4 & \multirow{2}{*}{$\kappa$} & $\textrm{diag}(1,1,1,0)$ & \multirow{2}{*}{$\textrm{diag}(0,0,0,1)$} & $\textrm{CSO}(3,0,1)$ \\[1mm]
\cline{1-1}\cline{3-3}\cline{5-5} 5 &  & $\textrm{diag}(1,1,-1,0)$ & &  $\textrm{CSO}(2,1,1)$\\[1mm]
\hline
\hline
6 & \multirow{5}{*}{$\kappa$} & 
\multirow{5}{*}{$\textrm{diag}(1,1,0,0)$} & $\textrm{diag}(0,0,1,1)$ & 
$\begin{array}{ll}\textrm{CSO}(2,0,2)\ , & \a\,\neq\,\frac{\pi}{4}\\[1mm]\mathfrak{f}_{1}\quad(\textrm{Solv}_{6})^{*} \ , & \a\,=\,\frac{\pi}{4}\end{array}$ \\[1mm]
\cline{1-1}\cline{4-4}\cline{5-5} 7 &  &  & $\textrm{diag}(0,0,1,-1)$ &  
$\begin{array}{ll}\textrm{CSO}(2,0,2)\ , & |\a|\,<\,\frac{\pi}{4}\\[1mm]\textrm{CSO}(1,1,2)\ , & |\a|\,>\,\frac{\pi}{4}\\[1mm]\mathfrak{g}_{0}\quad(\textrm{Solv}_{6})^{*} \ , & |\a|\,=\,\frac{\pi}{4}\end{array}$ \\[1mm]
\cline{1-1}\cline{4-4}\cline{5-5} 8 &  &  & $\textrm{diag}(0,0,0,1)$ &  $\mathfrak{h}_{1}\quad(\textrm{Solv}_{6})^{*}$ \\[1mm]
\hline
9 & \multirow{2}{*}{$\kappa$} & 
\multirow{2}{*}{$\textrm{diag}(1,-1,0,0)$} & $\textrm{diag}(0,0,1,-1)$ & 
$\begin{array}{ll}\textrm{CSO}(1,1,2)\ , & \a\,\neq\,\frac{\pi}{4}\\[1mm]\mathfrak{f}_{2}\quad(\textrm{Solv}_{6})^{*} \ , & \a\,=\,\frac{\pi}{4}\end{array}$\\[1mm]
\cline{1-1}\cline{4-4}\cline{5-5} 10 &  &  & $\textrm{diag}(0,0,0,1)$ &  $\mathfrak{h}_{2}\quad(\textrm{Solv}_{6})^{*}$ \\[1mm]
\hline\hline
11 & $\kappa$ & $\textrm{diag}(1,0,0,0)$ & $\textrm{diag}(0,0,0,1)$ &  
$\begin{array}{ll}\mathfrak{l}\,\,\,\,(\textrm{Nil}_{6}(3))^{*}\ , & \a\,\neq\,0\\[1mm]\textrm{CSO}(1,0,3) \ , & \a\,=\,0\end{array}$ \\[1mm]
\hline
\end{tabular}
}
\end{center}
\caption{{\it All the T-duality orbits of consistent deformations in the ${\bf 1} \, \oplus \, {\bf 10} \, \oplus \, {\bf 10^{\prime}}$ of half-maximal supergravity in $D=7$. Any value of $\,\alpha\,$ parameterises inequivalent orbits; the range of $\alpha$ is everywhere $-\frac{\pi}{2}\,<\,\alpha\,<\,\frac{\pi}{2}$, except in orbits 1, 2, 3 and 11, where it is reduced to $-\frac{\pi}{4}\,<\,\alpha\,\le\,\frac{\pi}{4}$ due to the symmetry w.r.t. interchanges between $Q$ \& $\tilde{Q}$.
Note that the value of $\kappa$, instead, can be restricted to being either $0$ or $1$ by using an $\mathbb{R}^{+}_{\Sigma}$ rescaling. For more details on algebras marked with *, see appendix~\ref{App:algebras}.} 
\label{table:orbits2}}
\end{table}

Please note that, whenever $\theta\,\neq\,0$, all the duality orbits belonging to this second branch of consistent deformations are related to orientifold compactifications of type II supergravities on a $\mathbb{T}^{3}$ with \emph{dyonic} generalised fluxes and hence go beyond those twisted reductions of DFT  considered in ref.~\cite{Dibitetto:2012rk}.

\section{Systematic analysis of critical points}
\label{sec:critical_points}

After having classified all the consistent deformations of half-maximal supergravity in $D=7$, the aim of this section is that of studying the critical points of the potential \eqref{VHalf_Max}. To this end we introduce the following explicit parametrisation for the $\textrm{SL}(4)$ scalars
\be
\label{vielbein}
{\mathcal{V}_{m}}^{\underline{m}} \ = \
\left(\begin{array}{cccc}
e^{\phi_{1}/2} & \chi_{1}\,e^{\phi_{2}/2} & \chi_{2}\,e^{\phi_{3}/2} & \chi_{4}\,e^{-(\phi_{1}+\phi_{2}+\phi_{3})/2} \\
0 & e^{\phi_{2}/2} & \chi_{3}\,e^{\phi_{3}/2} & \chi_{5}\,e^{-(\phi_{1}+\phi_{2}+\phi_{3})/2} \\
0 & 0 & e^{\phi_{3}/2} & \chi_{6}\,e^{-(\phi_{1}+\phi_{2}+\phi_{3})/2} \\
0 & 0 & 0 & e^{-(\phi_{1}+\phi_{2}+\phi_{3})/2}
\end{array}\right) \ ,
\ee
containing three dilatons and six axions. Please note that the vielbein in \eqref{vielbein} can be related to the object appearing in table~\ref{Table:dofs} in the following way
\be
{\mathcal{V}_{m}}^{\a\hat{\b}} \ = \ \frac{1}{\sqrt{2}} \, {\mathcal{V}_{m}}^{\underline{m}} \, \left[\Gamma_{\underline{m}}\right]^{\a\hat{\b}} \ ,
\ee
by using the Dirac matrices of $\textrm{SO}(4)$.

In terms of the vielbein ${\mathcal{V}_{m}}^{\underline{m}}$, the coset representative $M_{mn}$ appearing in the scalar potential is given by
\be
M_{mn} \ = \ {\mathcal{V}_{m}}^{\underline{m}} \, {\mathcal{V}_{n}}^{\underline{n}} \, \delta_{\underline{m}\underline{n}} \ ,
\ee
where $\delta_{\underline{m}\underline{n}}$ represents the $\textrm{SO}(4)$ invariant metric. By plugging the above parametrisation into the kinetic Lagrangian given in \eqref{Lkin}, one can rewrite it
as
\be
\label{Lkin_explicit}
\mathcal{L}_{\textrm{kin}} \ = \ -\frac{1}{2} \, K_{IJ}\,\left(\partial\Phi^{I}\right)\,\left(\partial\Phi^{J}\right) \ ,
\ee
where $\Phi^{I}\, \equiv \, \left(\Sigma,\,\phi_{1},\,\phi_{2},\,\phi_{3},\,\chi_{1},\,\chi_{2},\,\chi_{3},\,\chi_{4},\,\chi_{5},\,\chi_{6}\right)$, with $I\,=\,1,\,\dots,\,10$ and the kinetic metric 
$K_{IJ}$ assumes the form
\be
K_{IJ} \ = \ \left(
\begin{array}{c|c}
K^{(1)} & \\[2mm]
\hline
 & K^{(2)}
\end{array}
\right) \ ,
\ee
where the two $5\times 5$ blocks of $K_{IJ}$ have in general a complicated field-dependent expression. However, in the origin of the scalar manifold, they explicitly read
\be
\begin{array}{lcl}
\left.K^{(1)}\right|_{\textrm{origin}} \ = \ \left(
\begin{array}{ccccc}
 5 & 0 & 0 & 0 & 0  \\
 0 & \frac{1}{2} & \frac{1}{4} & \frac{1}{4} & 0 \\
 0 & \frac{1}{4} & \frac{1}{2} & \frac{1}{4} & 0 \\
 0 & \frac{1}{4} & \frac{1}{4} & \frac{1}{2} & 0 \\
 0 & 0 & 0 & 0 & \frac{1}{2}  
\end{array}\right) \ , & \textrm{and} &
\left.K^{(2)}\right|_{\textrm{origin}} \ = \ \frac{1}{2} \, \mathds{1}_{5} \ .
\end{array}
\ee

The inverse of the above kinetic metric turns out to be needed in order to write down the correct physical normalised mass matrix for $\Phi^{I}$, which then reads
\be
{\left(m^{2}\right)_{I}}^{J} \ \equiv \ \frac{1}{|V|}\,K^{JK}\,\partial_{K}\partial_{I}V \ .
\ee
We remind the reader that one needs to take the Breitenlohner-Freedman (BF) bound \cite{Breitenlohner:1982jf} into account when it comes to judging the stability of an AdS critical point. In $D$ dimensions, this translates into the following lower bound for the normalised mass of the mode in question
\be
\label{BF}
\frac{m^{2}}{|\Lambda|} \ \overset{!}{\geq} \ - \, \frac{D-1}{2\,(D-2)} \ ,
\ee
which equals $-\frac{3}{5}$ in 7D.

For the analysis of critical points of the scalar potential we will adopt the \emph{going to the origin} (GTTO) approach \cite{Dibitetto:2011gm}, \emph{i.e.} we will make use of a non-compact $\mathbb{R}^{+}\times\textrm{SL}(4)$ transformation in order to restrict the search of solutions to the origin of moduli space without loss of generality. 

Furthermore, since in both branches (1 \& 2) we retain a set of embedding tensor components which happens to be closed w.r.t. compact global symmetries as well, we are still allowed to use an $\textrm{SO}(4)$ to further simplify the embedding tensor while keeping all the scalars in the origin. In our case, we will exploit this possibility in order to assume a diagonal form for the symmetric matrix $Q_{mn}$.  

\subsection*{No-go argument for $\Lambda\neq 0$ within branch 1}

When $\theta\,=\,0$, all the non-vanishing embedding tensor irrep's happen to have the same $\mathbb{R}^{+}_{\Sigma}$ weight. As a consequence, the complete scalar potential within this class of deformations can be written as
\be
V(\Sigma,\,M_{mn}) \ = \ \Sigma^{-2} \, V_{0}(M_{mn}) \ ,
\ee
where $V_{0}$ is an arbitrary function of the $\textrm{SL}(4)$ scalars but independent of $\Sigma$.

This immediately implies that $\Sigma$ is generically a \emph{run-away} direction. This statement is analogous to that in ref.~\cite{deRoo:1985jh} concerning the \emph{run-away} behaviour of the $\textrm{SL}(2)$ dilaton in every purely electric gauging within half-maximal supergravity in four dimensions.

The only way of solving the $\Sigma$ field equation is having a vanishing $\Lambda$ at the solution. As a consequence, one is only left with Minkowski solutions of the no-scale type as the only possibility. 
An example of such a solution with $\xi_{mn}\,=\,\tilde{Q}^{mn}\,=\,0$ is 
\be
Q_{mn} \ = \ \left(\begin{array}{c|c} \mathds{1}_{2} & \\[2mm] \hline & 0_{2}\end{array}\right) \ ,
\ee
with the following (non-)normalised mass spectrum
\be
\begin{array}{lcccclcc}
\frac{1}{16} \quad (\times 2) & & , & & & 0 \quad (\times 8) & & .
\end{array}
\ee
Such a solution corresponds to a reduction of type I supergravity on an $\textrm{ISO}(2)$ group manifold \cite{Scherk:1979zr}. We have not explored this branch exhaustively but so far we have no evidence 
for the existence of Minkowski solutions with non-zero $\xi_{mn}$.

\subsection*{Critical points in branch 2}

In this second branch of consistent deformed theories $\theta$ offers us the only terms in the scalar potential having a different scaling behaviour w.r.t. $\Sigma$, thus allowing us to stabilise all the moduli at non-vanishing values of the cosmological constant.
This case represents the 7D analog of introducing non-trivial de Roo-Wagemans phases. 

In this branch, the QC \eqref{QCHalf_Max} take the following simple form
\be
\tilde{Q}^{mp}\,Q_{pn} \ - \ \frac{1}{4}\,\left(\tilde{Q}^{pq}\,Q_{pq}\right)\,\delta^{m}_{n} \ = \ 0 \ . 
\ee
If one furthermore restricts, as argued earlier to a diagonal $Q_{mn}$, the above QC imply a diagonal form for $\tilde{Q}^{mn}$ as well. This, in turn, guarantees that all the equations of motion for the axions will be automatically satisfied, thus simplifying our analysis enormously.

\subsubsection*{Exhaustive search within non-semisimple gaugings}

Within this class of theories there exist no-scale type Minkowski (\emph{i.e.} stable up to flat directions) and AdS solutions. These critical points are collected in table~\ref{table:Mkw} and \ref{table:AdS1}, respectively.  
\begin{table}[h!]
\begin{center}
\scalebox{1}[1]{
\begin{tabular}{| c | c | c | c | c | c |}
\hline
\textrm{ID} & $\theta$  & $Q_{mn}$ & $\tilde{Q}^{mn}$ & orbit & mass spectrum \\[1mm]
\hline \hline
1 & $0$ & $\textrm{diag}(\l,\l,0,0)$ & $\textrm{diag}(0,0,\m,\m)$ & 6 & 
$\begin{array}{cc}0 & (\times\,6)\\[1mm] \l^{2} & (\times\,2)\\[1mm]\m^{2} & (\times\,2)\end{array}$\\[1mm]
\hline
2 & $0$ & $\textrm{diag}(\l,\l,0,0)$ & $0_{4}$ & 6 & 
$\begin{array}{cc}0 & (\times\,8)\\[1mm] \frac{1}{16}\l^{2} & (\times\,2)\end{array}$\\[1mm]
\hline
\hline
3 & $\frac{\l}{2}$ & $\textrm{diag}(\l,0,0,0)$ & $0_{4}$ & 11 & 
$\begin{array}{cc}0 & (\times\,9)\\[1mm] \frac{1}{16}\l^{2} & (\times\,1)\end{array}$\\[1mm]
\hline
\end{tabular}
}
\end{center}
\caption{{\it All the Minkowski solutions of half-maximal supergravity in $D=7$ with non-semisimple gauge groups within branch 2. Please note that, in this case, the mass spectrum cannot be normalised w.r.t. to the value of the cosmological constant.} 
\label{table:Mkw}}
\end{table}
\begin{table}[h!]
\begin{center}
\scalebox{1}[1]{
\begin{tabular}{| c | c | c | c | c | c |}
\hline
\textrm{ID} & $\theta$  & $Q_{mn}$ & $\tilde{Q}^{mn}$ & orbit & mass spectrum \\[1mm]
\hline \hline
1 & $\frac{\l}{4}$ & $\textrm{diag}(\l,\l,\l,0)$ & $\textrm{diag}(0,0,0,\l)$ & 4 & 
$\begin{array}{cc}0 & (\times\,3)\\[1mm] -\frac{8}{15} & (\times\,1)\\[1mm] 
\frac{16}{15} & (\times\,5)\\[1mm] \frac{8}{3} & (\times\,1) \end{array}$\\[1mm]
\hline
2 & $\frac{\l}{14}$ & $\textrm{diag}(\l,\l,\l,0)$ & $\textrm{diag}(0,0,0,-\frac{8}{7}\l)$ & 4 & 
$\begin{array}{cc}0 & (\times\,3)\\[1mm] \frac{12}{5} & (\times\,5)\\[1mm] 
\frac{2}{35}\,\left(22\,\pm\,\sqrt{1954}\right) & (\times\,1)\end{array}$\\[1mm]
\hline
3 & $\frac{\l}{2}$ & $\textrm{diag}(\l,\l,\l,0)$ & $\textrm{diag}(0,0,0,\l)$ & 4 & 
$\begin{array}{cc}0 & (\times\,8)\\[1mm] \frac{4}{5} & (\times\,1)\\[1mm] 
\frac{12}{5} & (\times\,1)\end{array}$\\[1mm]
\hline
\end{tabular}
}
\end{center}
\caption{{\it All the AdS solutions of half-maximal supergravity in $D=7$ with non-semisimple gauge groups within branch 2. Sol.~1 is supersymmetric, whereas 2 \& 3 are non-supersymmetric. Sol.~2 
even violates the BF in \eqref{BF}, thus being unstable.} 
\label{table:AdS1}}
\end{table}

\subsubsection*{Exhaustive search within semisimple gaugings with $\tilde{Q}^{mn}\,=\,0$}

When considering the case of semisimple gaugings purely in the $\textbf{10}^{\prime}$, one has access to orbits 1, 2 and 3 of table~\ref{table:orbits2} with $\a\,=\,0$. These gaugings all admit an uplift to semisimple gaugings of the maximal theory where the embedding tensor is purely restricted to a $Y_{MN}\,\in\,\textbf{15}^{\prime}$ of $\textrm{SL}(5)$ of the form
\be
Y_{MN} \ = \ \left(
\begin{array}{c|c}
\theta & \\[1mm]
\hline
 & \frac{1}{2}\,Q_{mn}
\end{array}
\right) \ ,
\ee
and hence all fall into the classification of ref.~\cite{Samtleben:2005bp}. All critical points of this type are collected in table~\ref{table:AdS2}.
\begin{table}[h!]
\begin{center}
\scalebox{1}[1]{
\begin{tabular}{| c | c | c | c | c | c |}
\hline
\textrm{ID} & $\theta$  & $Q_{mn}$ & orbit & mass spectrum \\[1mm]
\hline \hline
1 & $\frac{\l}{2}$ & $\l\,\mathds{1}_{4}$ & 1 & 
$\begin{array}{cc}-\frac{8}{15} & (\times\,10)\end{array}$\\[1mm]
\hline
2 & $\l$ & $\l\,\mathds{1}_{4}$ & 1 & 
$\begin{array}{cc}-\frac{4}{5} & (\times\,9)\\[1mm] \frac{4}{5} & (\times\,1)\end{array}$\\[1mm]
\hline
3 & $\frac{\l}{2}$ & $\textrm{diag}(\l,\l,\l,2\l)$ & 1 & 
$\begin{array}{cc}0 & (\times\,3)\\[1mm] -\frac{4}{5} & (\times\,6)\\[1mm] \frac{4}{5} & (\times\,1)\end{array}$\\[1mm]
\hline
\end{tabular}
}
\end{center}
\caption{{\it All the AdS solutions of half-maximal supergravity in $D=7$ with semisimple gauge groups in the $\textbf{10}^{\prime}$. All these critical points admit an uplift to the maximal 7D theory, since they satisfy $\,\,\theta\,\tilde{Q}^{mn}\,=\,0$.
Sol.~1 is supersymmetric, whereas sol.~2 \& 3 are non supersymmetric and unstable.} 
\label{table:AdS2}}
\end{table}

\subsubsection*{Exploring semisimple gaugings in the $\textbf{10}\,\oplus\,\textbf{10}^{\prime}$}

When considering more general semisimple gaugings with both $Q$ \& $\tilde{Q}$ turned on, the space of solutions to the QC and field equations suddenly becomes much richer and a complete analytical treatment gets much harder to perform. However, we were able to exhaustively explore some particularly relevant subcases within this class.

By setting, \emph{e.g.} $q_{11}\,=\,1$ and $q_{33}\,=\,q_{44}$, one finds a set of isolated AdS/Minkowski solutions corresponding to critical points of $\textrm{SO}(4)$-gauged theories and two continuous branches with gauge group $\textrm{SO}(3,1)$ exhibiting an AdS -- Mkw -- dS transition and even allowing for a stable dS window.

The set of isolated solutions is presented in table~\ref{table:isolated}. In solutions 3 \& 4 in the table, the gauge group degenerates into $\textrm{ISO}(3)$ when $\l\,=\,0$. There one has a Minkowski critical point and then, when moving further into the $\l\,<\,0$ region, the gauge group becomes $\textrm{SO}(3,1)$ but, as one can see from the analytical $\l$-dependence of $V_{0}$ shown in the table, the cosmological constant goes back to negative instead of flipping sign. A similar transition has been first observed in ref.~\cite{Borghese:2012qm} in the context of the $\textrm{G}_{2}$ invariant sector of $\mathcal{N}=8$ supergravity in four dimensions.

In solution 5 instead, the gauge group evolves from $\textrm{SO}(4)$ to $\textrm{SO}(2,2)$ via $\textrm{CSO}(2,0,2)$ in correspondence of $\l\,=\,1$. However in this case, the cosmological constant stays vanishing for any value of $\l$.
\begin{table}[h!]
\begin{center}
\scalebox{1}[1]{
\begin{tabular}{| c | c | c | c | c | c | c |}
\hline
\textrm{ID} & $\theta$  & $Q_{mn}$ & $\tilde{Q}^{mn}$ & orbit & $V_0$ & mass spectrum \\[1mm]
\hline \hline
1 & $\frac{1-\l}{2}$ & $\mathds{1}_{4}$ & $\l\,\mathds{1}_{4}$ & 1 & $-\frac{15}{4}\,(1-\l)^{2}$ &
$\begin{array}{cc}-\frac{8}{15} & (\times\,10)\end{array}$\\[1mm]
\hline
2 & $\frac{1-\l}{4}$ & $\textrm{diag}(1,1,1,\l)$ & $\textrm{diag}(\l,\l,\l,1)$ & 1 - 4 - 2 & $-\frac{15}{16}\,(1-\l)^{2}$ &
$\begin{array}{cc}0 & (\times\,3)\\[1mm] -\frac{8}{15} & (\times\,1) \\[1mm] \frac{16}{15} & (\times\,5)
\\[1mm] \frac{8}{3} & (\times\,1)\end{array}$\\[1mm]
\hline
3 & $1-\l$ & $\mathds{1}_{4}$ & $\l\,\mathds{1}_{4}$ & 1 & $-5\,(1-\l)^{2}$ &
$\begin{array}{cc}-\frac{4}{5} & (\times\,9)\\[1mm]\frac{4}{5} & (\times\,1)\end{array}$\\[1mm]
\hline
4 & $\frac{1-\l}{2}$ & $\textrm{diag}(1,1,1,\l)$ & $\textrm{diag}(\l,\l,\l,1)$ & 1 - 4 - 2 & $-\frac{5}{4}\,(1-\l)^{2}$ &
$\begin{array}{cc}0 & (\times\,8)\\[1mm] \frac{4}{5} & (\times\,1) \\[1mm] \frac{12}{15} & (\times\,1)\end{array}$\\[1mm]
\hline\hline
5 & $0$ & $\textrm{diag}(1,1,\l,\l)$ & $\textrm{diag}(\l,\l,1,1)$ & 1 - 6 - 3 & $0$ &
$\begin{array}{cc}0 & (\times\,6)\\[1mm] 4\,(1-\l)^{2} & (\times\,4)\end{array}$\\[1mm]
\hline
\end{tabular}
}
\end{center}
\caption{{\it Isolated AdS/Mkw solutions of half-maximal supergravity in $D=7$ with semisimple gauge groups in the $\textbf{10}\,\oplus\,\textbf{10}^{\prime}$. Note that the gauge group transitions in rows 3, 4 and 5 do not involve a sign change for $V_{0}$.
Sol.~1 \& 2 are supersymmetric, whereas all the other solutions happen to break supersymmetry. Sol.~3 is the only one violating the BF bound \eqref{BF}.}\label{table:isolated}}
\end{table}    

\subsubsection*{Stable dS in the $\textrm{SO}(3,1)$-gauged theory (orbit 3)}

\begin{figure}[h!]
\begin{center}
\begin{tabular}{cc}
\includegraphics[scale=0.7,keepaspectratio=true]{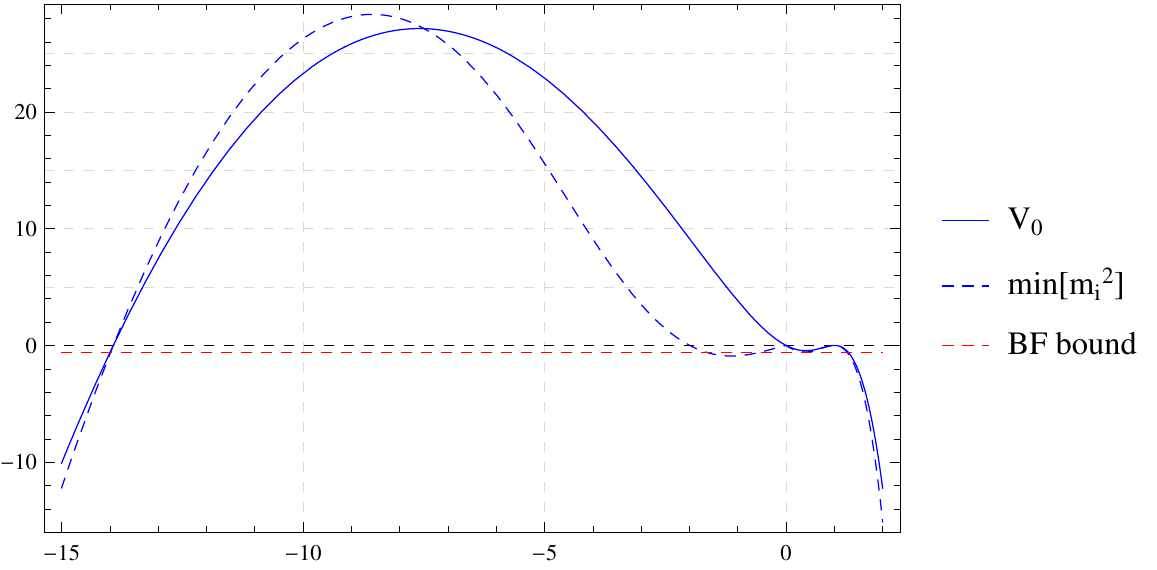}  &  \includegraphics[scale=0.7,keepaspectratio=true]{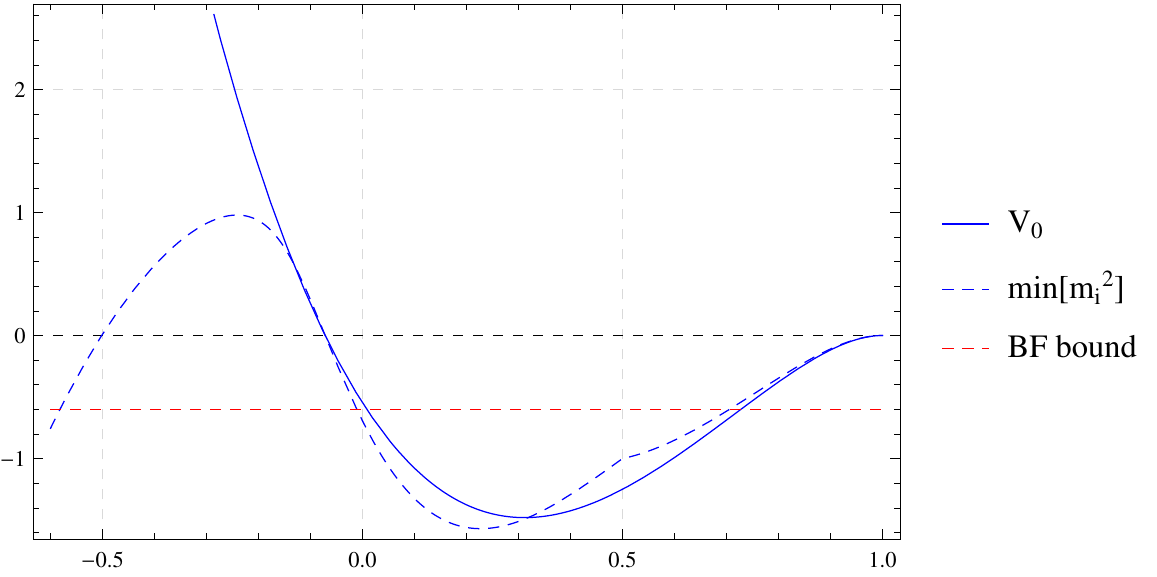}
\end{tabular}
\caption{{\it The value of the cosmological constant ($V_{0}$) and the minimum (non-vanishing) eigenvalue of the normalised mass matrix as a function of the $\lambda$ parameter. The plot on the left represents the
window of stable dS for the solution labelled by ``$+$'', whereas the right plot shows the one for the solution labelled by ``$-$''.}}
\label{dS_plots}
\end{center}
\end{figure}

The continuous branches of solutions (labelled by $\pm$) read
\be
\begin{array}{lccclccclc}
Q_{\pm} \ = \ \textrm{diag}(1,\l,\l,\l) & & , & & \tilde{Q}_{\pm} \ = \ f_{\pm}(\l)\,\textrm{diag}(\l,1,1,1) & & \textrm{and} & & \theta_{\pm} \ = \ g_{\pm}(\l) & ,
\end{array} \notag
\ee
where
\be
f_{\pm}(\l) \ \equiv \ \frac{-7+22\l-7\l^{2}\,\pm\,(1-\l)\sqrt{49-82\l+49\l^{2}}}{8\,(2-\l)} \ \ ,
\notag
\ee
and
\be
g_{\pm}(\l) \ \equiv \ \left(\frac{1}{1-\l}\,+\,\frac{15}{8+8\l\,\pm\,\sqrt{49-82\l+49\l^{2}}}\right)^{-1} \ \ .
\notag
\ee
The above solutions exhibit a \emph{stable dS} window respectively given by
\be
\begin{array}{lccclccclcccc}
-7-4\sqrt{3} & < & \l & < & \m_{+} & & \textrm{and} & & \m_{-} & < & \l & < & -7+4\sqrt{3} \ ,
\end{array}
\notag
\ee
where $\m_{\pm}$ represent the two real roots\footnote{Numerically, $\m_{\pm}\,=\,\frac{1}{56} \left(-11-3 \sqrt{385}\,\pm\,\sqrt{450+66 \sqrt{385}}\right)$.} of the following polynomial ($\m_{+}\,<\,\m_{-}$)
\be
P(\m) \ \equiv \ 98 \,+\,77\m\,-\,222\m^{2}\,+\,77\m^{3}\,+\,98\m^{4} \ = \ 0 \ .
\ee
This situation is depicted in figure~\ref{dS_plots}.
Such stable dS windows lie in the vicinity of a stable Minkowski critical point in analogy to what has been observed in refs~\cite{deCarlos:2009qm,Danielsson:2012by,Blaback:2013ht} in the context of $\mathcal{N}=1$ supergravity in 4D.

In the continuous branch labelled by $+$, one is approaching the stable Minkowski solution at $\l\,=\,-7-4\sqrt{3}$
\be
\begin{array}{lccclc}
Q_{+} \ = \ \textrm{diag}(1,-7-4\sqrt{3},-7-4\sqrt{3},-7-4\sqrt{3}) & & , & & \tilde{Q}_{+} \ = \ \textrm{diag}(7+4\sqrt{3},-1,-1,-1) &  ,
\end{array} \notag
\ee
and $\theta_{+} \ = \ 0\,$, where the non-normalised mass spectrum reads
\be
\begin{array}{cccccccc}
0 & (\times\,4) & & , & & 32\,(7+4\sqrt{3}) & (\times\,6) & . 
\end{array} \notag
\ee

An explicit example of stable dS critical point within branch $+$ ($\l\,=\,-3$) is given by
\be
\begin{array}{lclclc}
Q_{+} \ = \ \textrm{diag}(1,-3,-3,-3) & , &  \tilde{Q}_{+} \ = \ \dfrac{-17+2\sqrt{46}}{5}\,\textrm{diag}(-3,1,1,1) &  \textrm{and} &  \theta_{+} \ = \ \dfrac{-6+\sqrt{46}}{5} & .
\end{array} \notag
\ee
In this case the value of the cosmological constant reads $V_{0}\,=\,\frac{16}{5}\,(52-7\sqrt{46})$, and the mass spectrum is
\be
\begin{array}{ccccccccc}
0 & (\times\,3) & , & \dfrac{28+\sqrt{46}}{15} & (\times\,5) & , & \dfrac{1}{90}\,\left(212-13\sqrt{46}\pm\sqrt{61310-7504\sqrt{46}}\right) & (\times\,1) & .
\end{array} \notag
\ee

This example of dS vacuum is highly non-geomeric. In the standard language of Neveu-Schwarz fluxes \cite{Shelton:2005cf}, it includes not only the three-form $H$-flux and the metric $\omega$-flux, but also the non-geometric $Q$ and $R$-fluxes. None of them can be removed through a duality transformation, and then the orbit to which they belong is genuinely non-geometric in the sense of \cite{Dibitetto:2012rk}. While the universal half-maximal gaugings were uplifted to duality covariant higher-dimensional theories in \cite{Aldazabal:2011nj,Geissbuhler:2011mx}, massive deformations have been considered in that context in \cite{Hohm:2011cp}. We believe that combining these results can provide an uplift of this vacuum eventually.

\section{Conclusions}
\label{sec:conclusions}

In this paper we have studied various aspects of supergravities in $D=7$ with sixteen supercharges coupled to three vector multiplets.
The most general deformations include a combination of a ``Romans-like'' massive deformation and a traditional gauging of a subgroup of the global duality group.

By using the embedding tensor formalism, we were first able to classify the inequivalent duality orbits of consistent deformations. It is worth mentioning that all orbits with no massive deformation
can be regarded as generalised twisted reductions of DFT, provided that one allows for a dependence on doubled coordinates generically violating the section condition. 
The above massive deformation happens to have a non-trivial S-duality phase. Hence, any attempt of uplifting those orbits where a gauging is combined with such a deformation would require going beyond 
DFT reductions.

Secondly, we studied the properties of the different scalar potentials induced by the aforementioned deformations when it comes to critical points. We found that all orbits of gauged theories without 
massive deformation can only admit no-scale type Minkowski solutions. On the contrary, when the massive deformation is turned on together with a gauging, various types of maximally symmetric solutions
appear, the zoology of such models including interesting examples of (non-)supersymmetric AdS and stable dS vacua. To our knowledge, this is the first example of a stable dS critical point obtained
through spontaneous supersymmetry breaking within a theory with such a large amount of supercharges.   

%
%

\section*{Acknowledgments}

The work of GD is supported by the Swedish Research Council (VR) and JJF-M is funded by the Fundaci\'on S\'eneca - Talento Investigador Program. GD \& JJF-M would like to thank the Instituto de Astronom\'{\i}a y F\'{\i}sica del Espacio (IAFE) in Buenos Aires for warm hospitality while part of this project was carried out.

\newpage

%
%

\appendix

\section*{Summary of indices}
All throughout the text we extensively make use of indices of different groups. Here we give a list of the notations retained in this work
\be
\begin{array}{ll}
\mathbb{M},\,\mathbb{N},\,\dots & \textrm{fundamental of } \, \textrm{SO}(6,6) \\[2mm]
A,\,B,\,\dots & \textrm{fundamental of } \, \textrm{SO}(3,3) \\[2mm]
M,\,N,\,\dots & \textrm{fundamental of } \, \textrm{SL}(5) \\[2mm]
m,\,n,\,\dots & \textrm{fundamental of } \, \textrm{SL}(4) \\[2mm]
\underline{m},\,\underline{n},\,\dots & \textrm{fundamental of } \, \textrm{SO}(4)_{\textrm{local}} \\[2mm]
a,\,b,\,\dots & \textrm{fundamental of } \, \textrm{SL}(2) \\[2mm]
\m,\,\n,\,\dots & \textrm{7D spacetime indices}  \\[2mm]
\a,\,\b,\,\dots & \textrm{fundamental of } \, \textrm{SU}(2)_{R} \\[2mm]
\hat{\a},\,\hat{\b},\,\dots & \textrm{fundamental of } \, \textrm{SU}(2) \\[2mm]
I,\,J,\,\dots & \textrm{collective labels for 7D scalars}  \\[2mm]
\end{array} \notag
\ee

\section{Non-semisimple gauge algebras}
\label{App:algebras}

In section~\ref{sec:orbits} we have studied the T-duality orbits of consistent deformations in half-maximal $D=7$ supergravity and for each of them, we identified the underlying gauge algebra and collected 
the results in tables~\ref{table:orbits1} \& \ref{table:orbits2}. 
Since there exists no exhaustive classification of non-semisimple algebras of dimension six, we would like to explicitly give the form of the algebras appearing in tables~\ref{table:orbits1} \& \ref{table:orbits2}.

\subsection*{Solvable algebras}

This class includes the gaugings described in rows 6 -- 11 of table~\ref{table:orbits1} and 6 -- 10 of table~\ref{table:orbits2}. In the former case, the central generator named $z$ will be realised
through $\mathbb{R}^{+}_{\Sigma}$.

\subsubsection*{The CSO($2,0,2$) and CSO($1,1,2$) algebras}

The details about these algebras can be found in ref.~\cite{deRoo:2006ms}; we summarise here some relevant facts.
The six generators are labelled as $\{t_{0},\,t_{i},\,s_{i},\,z\}_{i=1,2}$, where $t_{0}$ generates SO($2$) (SO($1,1$)), under which $\{t_{i}\}$ and $\{s_{i}\}$ transform as doublets
\be
\begin{array}{cccc}
\left[t_{0},\,t_{i}\right]\,=\,{\epsilon_{i}}^{j}\,t_{j} & , &
\left[t_{0},\,s_{i}\right]\,=\,{\epsilon_{i}}^{j}\,s_{j} & ,
\end{array}
\ee
where the Levi-Civita symbol ${\epsilon_{i}}^{j}$ has one index lowered with the metric $\eta_{ij}\,=\,$diag$(\pm 1,1)$ depending on the two different signatures. $z$ is a central charge appearing in the following commutators
\be 
\left[t_{i},\,s_{j}\right]\,=\,\eta_{ij}\,z\ . 
\ee
The Cartan-Killing metric is diag($\mp 1, \underbrace{0, \cdots, 0}_{\textrm{5 times}}$), where the $\mp$ is again related to the two different signatures.

\subsubsection*{The $\mathfrak{f}_{1}$ and $\mathfrak{f}_{2}$ algebras}

These are of the form Solv$_{4}\,\times\,$U$(1)^{2}$. The 4 generators of Solv$_{4}$ are labeled by $\{t_{0},\,t_{i},\,z\}_{i=1,2}$, where $t_{0}$ generates SO($2$) (SO($1,1$)), under which $\{t_{i}\}$ transform as a doublet
\be 
\left[t_{0},\,t_{i}\right]\,=\,{\epsilon_{i}}^{j}\,t_{j}\ , 
\ee
\be 
\left[t_{i},\,t_{j}\right]\,=\,\epsilon_{ij}\,z\ . 
\ee
The Cartan-Killing metric is diag($\mp 1, \underbrace{0, \cdots, 0}_{\textrm{5 times}}$).

\subsubsection*{The $\mathfrak{h}_{1}$ and $\mathfrak{h}_{2}$ algebras}

The 6 generators are $\{t_{0},\,t_{i},\,s_{i},\,z\}_{i=1,2}$ and they satisfy the following commutation relations
\be
\begin{array}{lclc}
\left[t_{0},\,t_{i}\right]\,=\,{\epsilon_{i}}^{j}\,t_{j} & , &
\left[t_{0},\,s_{i}\right]\,=\,{\epsilon_{i}}^{j}\,s_{j}\,+\,t_{i} &
, \\[2mm]
\left[t_{i},\,s_{j}\right]\,=\,\eta_{ij}\,z & , &
\left[s_{i},\,s_{j}\right]\,=\,\epsilon_{ij}\,z & .
\end{array}
\ee
The Cartan-Killing metric is diag($\mp 1, \underbrace{0, \cdots, 0}_{\textrm{5 times}}$).

\subsubsection*{The $\mathfrak{g}_{0}$ algebra}

The 6 generators are $\{t_{0},\,t_{I},\,z\}_{I=1,\cdots,4}$, where $t_{0}$ transforms cyclically the $\{t_{I}\}$ amongst themselves such that
\be
\left[\bigg[\big[[t_{I},\,t_{0}],\,t_{0}\big],\,t_{0}\bigg],\,t_{0}\right]\,=\,t_{I}\ , 
\ee
and
\be 
\left[t_{1},\,t_{3}\right]\,=\,\left[t_{2},\,t_{4}\right]\,=\,z\ .
\ee
Note that this algebra is solvable and not nilpotent even though its Cartan-Killing metric is \emph{completely zero}.

\subsection*{Nilpotent algebras}

This family comprises the gaugings found in rows 12 and 11 in tables~\ref{table:orbits1} and \ref{table:orbits2}, respectively.

\subsubsection*{The CSO($1,0,3$) algebra}

The details about this algebra can be again found in ref.~\cite{deRoo:2006ms}; briefly summarising, the 6 generators are given by  $\{t_{m},\,z^{m}\}_{m=1,2,3}$ and they satisfy the following commutation relations
\be 
\left[t_{m},\,t_{n}\right]\,=\,\epsilon_{mnp}\,z^{p}\ ,
\ee
with all the other brackets being vanishing. The order of nilpotency of this algebra is 2.

\subsubsection*{The $\mathfrak{l}$ algebra}

The 6 generators $\{t_{1},\cdots,\,t_{6}\}$ satisfy the following commutation relations
\be
\begin{array}{lclclc}
 \left[t_{1},\,t_{2}\right]\,=\,t_{4} & , &
 \left[t_{1},\,t_{4}\right]\,=\,t_{5} & , &
 \left[t_{2},\,t_{4}\right]\,=\,t_{6} & .
\end{array}
\ee
The corresponding central series reads
\be
\begin{array}{lclclcl}
 \left\{t_{1},\,t_{2},\,t_{3},\,t_{4},\,t_{5},\,t_{6}\right\} &\supset&
 \left\{t_{4},\,t_{5},\,t_{6}\right\} &\supset&
 \left\{t_{5},\,t_{6}\right\} &\supset& \left\{0\right\}\ ,
\end{array}
\ee
from which we can immediately conclude that its nilpotency order is 3.

\section{Mapping between $\textrm{SL}(4)$ and $\textrm{SO}(3,3)$}
\label{App:'tHooft}

The 't Hooft symbols $\left[G_{A}\right]^{mn}$ are invariant tensors which map the fundamental representation of $\textrm{SO}(3,3)$, \emph{i.e.} the \textbf{6} into the anti-symmetric two-form of $\textrm{SL}(4)$
\be
v^{mn} \ = \ \left[G_{A}\right]^{mn} \, v^{A} \ ,
\label{mn2A}
\ee
for any object where $v^{A}$ transforming as a vector of $\textrm{SO}(3,3)$. The two-form irrep of $\textrm{SL}(4)$ is real due to the role of the Levi-Civita tensor relating $v_{mn}$ to $v^{mn}$ via
\be
v_{mn} \ = \ \frac{1}{2} \, \epsilon_{mnpq} \, v^{pq} \ .
\ee

The inverse of the mapping in \eqref{mn2A} is carried out by the corresponding 't Hooft symbols with lower indices, \emph{i.e.} $\left[G_{A}\right]_{mn} \, \equiv \, \frac{1}{2} \, \epsilon_{mnpq} \, \left[G_{A}\right]^{pq}$. The tensors $\left[G_{A}\right]^{mn}$ and $\left[G_{A}\right]_{mn}$ satisfy the following identities
\be
\begin{array}{lc}
\left[G_{A}\right]_{mn} \, \left[G_{B}\right]^{mn} \ = \ 2\,\eta_{AB} & , \\[2mm]
\left[G_{A}\right]_{mp} \, \left[G_{B}\right]^{pn} \, + \, \left[G_{B}\right]_{mp} \, \left[G_{A}\right]^{pn} \ = \ -\d^{n}_{m}\,\eta_{AB} & , \\[2mm]
\left[G_{A}\right]_{mp} \, \left[G_{B}\right]^{pq} \, \left[G_{C}\right]_{qr} \, \left[G_{D}\right]^{rs} \, \left[G_{E}\right]_{st} \, \left[G_{F}\right]^{tn}\ = \ \d^{n}_{m}\,\epsilon_{ABCDEF} & , 
\end{array}
\ee
where $\eta_{AB}$ and $\epsilon_{ABCDEF}$ denote the $\textrm{SO}(3,3)$ light-cone metric and the Levi-Civita symbol, respectively. 

We adopt the following explicit representation for the 't Hooft symbols in light-cone coordinates
\be
\label{tHooft}
\begin{array}{lclc}
\left[G_{1}\right]^{mn} \ = \ \left(
\begin{array}{cccc}
0 & -1 & 0 & 0 \\
1 & 0 & 0 & 0 \\
0 & 0 & 0 & 0 \\
0 & 0 & 0 & 0 
\end{array}\right)  & , &
\left[G_{\bar 1}\right]^{mn} \ = \ \left(
\begin{array}{cccc}
0 & 0 & 0 & 0 \\
0 & 0 & 0 & 0 \\
0 & 0 & 0 & -1 \\
0 & 0 & 1 & 0 
\end{array}\right) & , \\[7mm]
\left[G_{2}\right]^{mn} \ = \ \left(
\begin{array}{cccc}
0 & 0 & -1 & 0 \\
0 & 0 & 0 & 0 \\
1 & 0 & 0 & 0 \\
0 & 0 & 0 & 0 
\end{array}\right)  & , &
\left[G_{\bar 2}\right]^{mn} \ = \ \left(
\begin{array}{cccc}
0 & 0 & 0 & 0 \\
0 & 0 & 0 & -1 \\
0 & 0 & 0 & 0 \\
0 & 1 & 0 & 0 
\end{array}\right) & , \\[7mm]
\left[G_{3}\right]^{mn} \ = \ \left(
\begin{array}{cccc}
0 & 0 & 0 & -1 \\
0 & 0 & 0 & 0 \\
0 & 0 & 0 & 0 \\
1 & 0 & 0 & 0 
\end{array}\right)  & , &
\left[G_{\bar 3}\right]^{mn} \ = \ \left(
\begin{array}{cccc}
0 & 0 & 0 & 0 \\
0 & 0 & -1 & 0 \\
0 & 1 & 0 & 0 \\
0 & 0 & 0 & 0 
\end{array}\right) & . 
\end{array}
\ee

\section{$\textrm{SO}(4)$ Dirac matrices in the Weyl representation}
\label{App:Gamma}

In $0+4$ dimensions Dirac spinors have $4$ complex components; however, such spinors are not irreducible. Every Dirac spinor splits into a pair of chiral (Weyl) spinors caarying $2$ independent complex components each.
We therefore choose the following Weyl representation for the Dirac matrices, \emph{i.e.} where they all assume the form 
\be
\Gamma_{\underline{m}} \ = \ \left(
\begin{array}{c|c}
0_{2} & \left[\Gamma_{\underline{m}}\right]^{\a\hat{\b}} \\
\hline
\left[\bar{\Gamma}_{\underline{m}}\right]_{\hat{\a}\b} & 0_{2}
\end{array}\right) \ ,
\ee
which needs to satisfy
\be
\left\{\Gamma_{\underline{m}},\,\Gamma_{\underline{n}}\right\} \ = \ 2\,\delta_{\underline{m}\underline{n}} \, \mathds{1}_{4} \ .
\ee

We perform the following explicit choice for the chiral $2\times 2$ blocks
\be
\begin{array}{lclc}
\Gamma_{\underline{1}} \ = \ \left(
\begin{array}{c|c}
0_{2} & \mathds{1}_{2} \\
\hline
\mathds{1}_{2} & 0_{2}
\end{array}\right)  & , &
\Gamma_{\underline{2}} \ = \ \left(
\begin{array}{c|c}
0_{2} & i\,\sigma^{1} \\
\hline
-i\,\sigma^{1} & 0_{2}
\end{array}\right) & , \\[5mm]
\Gamma_{\underline{3}} \ = \ \left(
\begin{array}{c|c}
0_{2} & i\,\sigma^{2}\\
\hline
-i\,\sigma^{2} & 0_{2}
\end{array}\right)  & , &
\Gamma_{\underline{4}} \ = \ \left(
\begin{array}{c|c}
0_{2} & i\,\sigma^{3} \\
\hline
-i\,\sigma^{3} & 0_{2}
\end{array}\right) & ,
\end{array}
\ee
where $\left\{\sigma^{i}\right\}_{i\,=\,1,\,2,\,3}$ are the usual Pauli matrices given by
\be
\begin{array}{lclclc}
\sigma^{1} \ = \ \left(
\begin{array}{cc}
0 & 1 \\
1 & 0
\end{array}\right)  & , &
\sigma^{2} \ = \ \left(
\begin{array}{cc}
0 & -i \\
i & 0
\end{array}\right)  & , &
\sigma^{3} \ = \ \left(
\begin{array}{cc}
1 & 0 \\
0 & -1
\end{array}\right) & .
\end{array}
\ee
%

%
%

\small

\bibliography{references}
\bibliographystyle{utphys}

\end{document}